\documentclass[11pt,twoside]{article}

\usepackage{asp2014}

\resetcounters

\aspvoltitle{ASTRONOMICAL DATA ANALYSIS SOFTWARE AND SYSTEMS XXVI}
\aspvolume{in press}
\aspcpryear{ }
\aspvolauthor{ }

\begin{document}

\title{A Command-line Cross-matching tool for modern astrophysical pipelines}

\author{Giuseppe~Riccio,$^1$ Massimo~Brescia,$^1$ Stefano~Cavuoti,$^1$ Amata~Mercurio,$^1$ Anna~Maria~Di~Giorgio,$^2$ and Sergio~Molinari$^2$
\affil{$^1$INAF OACN - Astronomical Observatory of Capodimonte, Napoli, Italy; \email{giuseppe.riccio08@gmail.com}}
\affil{$^2$INAF IAPS - Istituto di Astrofisica e Planetologia Spaziali, Roma, Italy}}

\paperauthor{Giuseppe~Riccio}{giuseppe.riccio08@gmail.com}{0000-0001-7020-1172}{INAF - Istituto Nazionale di Astrofisica}{OACN - Astronomical Observatory of Capodimonte}{Napoli}{Napoli}{80131}{Italy}
\paperauthor{Massimo~Brescia}{brescia@oacn.inaf.it}{}{INAF - Istituto Nazionale di Astrofisica}{OACN - Astronomical Observatory of Capodimonte}{Napoli}{Napoli}{80131}{Italy}
\paperauthor{Stefano~Cavuoti}{stefano.cavuoti@gmail.com}{}{INAF - Istituto Nazionale di Astrofisica}{OACN - Astronomical Observatory of Capodimonte}{Napoli}{Napoli}{80131}{Italy}
\paperauthor{Amata~Mercurio}{mercurio@na.astro.it}{}{INAF - Istituto Nazionale di Astrofisica}{OACN - Astronomical Observatory of Capodimonte}{Napoli}{Napoli}{80131}{Italy}
\paperauthor{Anna~Maria~Di~Giorgio}{Anna.DiGiorgio@iaps.inaf.it}{}{INAF - Istituto Nazionale di Astrofisica}{IAPS - Istituto di Astrofisica e Planetologia Spaziali}{Roma}{Roma}{00133}{Italy}
\paperauthor{Sergio~Molinari}{molinari@iaps.inaf.it}{}{INAF - Istituto Nazionale di Astrofisica}{IAPS - Istituto di Astrofisica e Planetologia Spaziali}{Roma}{Roma}{00133}{Italy}

\begin{abstract}
The emerging need for efficient, reliable and scalable astronomical catalog cross-matching is becoming more pressing in the current data-driven science era, where the size of data has rapidly increased up to the Petabyte scale. $C^3$ (\textit{Command-line Catalogue Cross-matching}) is a multi-platform tool designed to efficiently cross-match massive catalogues from modern astronomical surveys, ensuring high-performance capabilities through the use of a multi-core parallel processing paradigm. The tool has been conceived to be executed as a stand-alone command-line process or integrated within any generic data reduction/analysis pipeline, providing the maximum flexibility to the end user, in terms of parameter configuration, coordinates and cross-matching types. In this work we present the architecture and the features of the tool. Moreover, since the modular design of the tool enables an easy customization to specific use cases and requirements, we present also an example of a customized $C^3$ version designed and used in the FP7 project ViaLactea, dedicated to cross-correlate Hi-GAL clumps with multi-band compact sources.
\end{abstract}

\section{Introduction}
The data volumes from the ongoing and next generation multi-band and multi-epoch surveys are expected to be so huge that analyzing, cross-correlating and extracting knowledge from such data will represent a challenge for scientists and computer engineers. Therefore, efficient techniques and software solutions, to be directly integrated into the reduction pipelines, will be required to cross-correlate in real time a large variety of parameters for billions of sky objects. One of the most common techniques used in astrophysics and a core step of any standard modern pipeline of data reduction/analysis, particularly sensible to the growing of the datasets dimensions, is the cross-match among heterogeneous catalogues, which consists in identifying and comparing sources belonging to different observations, performed at different wavelengths or under different conditions.

In this work we present $C^{3}$ (\textit{Command-line Catalogue Cross-match}), a tool to perform efficient catalogue cross-matching, based on the multi-thread paradigm, which can be easily integrated into an automatic data analysis pipeline. Furthermore, one of major features of this tool is the possibility to choose shape, orientation and size of the cross-matching area, making the $C^{3}$ tool easily tailorable on the specific user needs.

\section{$C^3$ Design and Features}
$C^3$ is a command-line software, designed and developed to perform general cross-matching among astrophysical catalogues, matching the needs to work on large datasets produced by independent surveys, to combine data to extract new information and to increase the astrophysical knowledge. Based on a specialized sky partitioning function, its high-performance capability is ensured by making use of the multi-core parallel processing paradigm. It works with the most common catalogue formats: FITS, ASCII, CSV, VOTable and with equatorial and galactic coordinates systems.

In order to be a general purpose tool, different functional cases have been implemented (see Sect.~\ref{sect:usecases} for further details), as well as the most used join function types (Sect.~\ref{sect:join}). However, $C^{3}$ is easy to use and to configure through few lines in a single configuration file. Main features of $C^{3}$ are the following:

\begin{itemize}
 \item \textit{Command line}: it can be used as stand-alone process or integrated within more complex pipelines;
 \item \textit{Python compatibility}: up to version 3.4;
 \item \textit{Multi-platform}: $C^{3}$ has been tested on Ubuntu Linux 14.04, Windows 7/10, Mac OS and Fedora;
 \item \textit{Multi-process}: the cross-matching process has been developed to run by using a multi-core parallel processing paradigm;
 \item \textit{Sky partitioning}: a simple sky partitioning algorithm is used to reduce computational time;
 \item \textit{User-friendliness}: the tool is very simple to configure and to use; only a simple configuration file is required.
\end{itemize}

\subsection{Functional cases}\label{sect:usecases}
$C^{3}$ is able to match two input catalogues by applying three different matching criteria, corresponding to different functional cases: \textit{Sky}, \textit{Exact Value} and \textit{Row-by-Row}. In the \textit{Sky} functional case, $C^{3}$ performs a positional cross-match between two catalogues, based on the same concept of the Q-FULLTREE approach, an our tool introduced in \citep{becciani2015}: for each object of the first input catalogue, it is possible to define an elliptical, circular (as special ellipse) or rectangular region centered on its coordinates, whose dimensions are limited by a fixed value or defined by specific catalogue parameters (for instance, the FWHM at a specific wavelength). The orientation in the space of the region is defined by a position angle, characterized also by two additional parameters, to opportunely set the zero-point and the direction of rotation. Once defined the region of interest, the next step is to search for sources of the second catalogue within such region, by comparing their distance from the central object and the limits of the area.



In the \textit{Exact Value} case, two objects are matched if they have the same value for a pair of columns (one for each catalogue) defined in the configuration file. Finally the \textit{Row-by-Row} case consists in matching objects with the same row-ID of the two catalogues.

\subsection{Match selection and join types}\label{sect:join}

The results of the cross-match are stored as a series of rows, corresponding to the matching objects. In the \textit{Exact value} and \textit{Sky} cases, the conditions that matched rows have to satisfy to be stored can be chosen between two match selection criteria (\textit{all} the matches or only the \textit{best} pairs) and a number of different join possibilities: \textit{i)} \textbf{$1$ and $2$}, only rows having an entry in both input catalogues; \textit{ii)} \textbf{$1$ or $2$}, all rows, matched and unmatched, from both input catalogues; \textit{iii)} \textbf{All from $1$ (All from $2$)}, all matched rows from catalogue $1$ (or $2$), together with the unmatched rows from catalogue $1$ (or $2$); \textit{iv)} \textbf{$1$ not $2$ ($2$ not $1$)}, all the rows of catalogue $1$ (or $2$) without matches in the catalogue $2$ (or $1$); \textit{v)} \textbf{$1$ xor $2$}, only unmatched rows from the catalogue $1$ and $2$.


\subsection{Performance boosters}

In order to increase the performance in terms of computational time, $C^3$ makes use of two different methods: \textit{i)} a massive application of multi-core parallel processing paradigm, managed by defining, in the configuration file, the number of concurrent parallel processes; \textit{ii)} a sky partitioning algorithm, in order to reduce the number of checks between the two catalogues. A subsample of the first input catalogue is assigned to each concurrent process, while the objects of the second catalogue are assigned to one of the \textit{cells} defined by the partitioning procedure, according to their coordinates. The size of the unit cell is defined by the maximum dimension that the matching regions can assume, (Fig.~\ref{fig:partitioning}a) . The described choice to set the dimensions of the cells ensures that the matching between an object of the first and a source of the second catalogue can happen only if the source lies in the  nine cells surrounding the object, (Fig.~\ref{fig:partitioning}b).


\articlefigure[width=.7\textwidth]{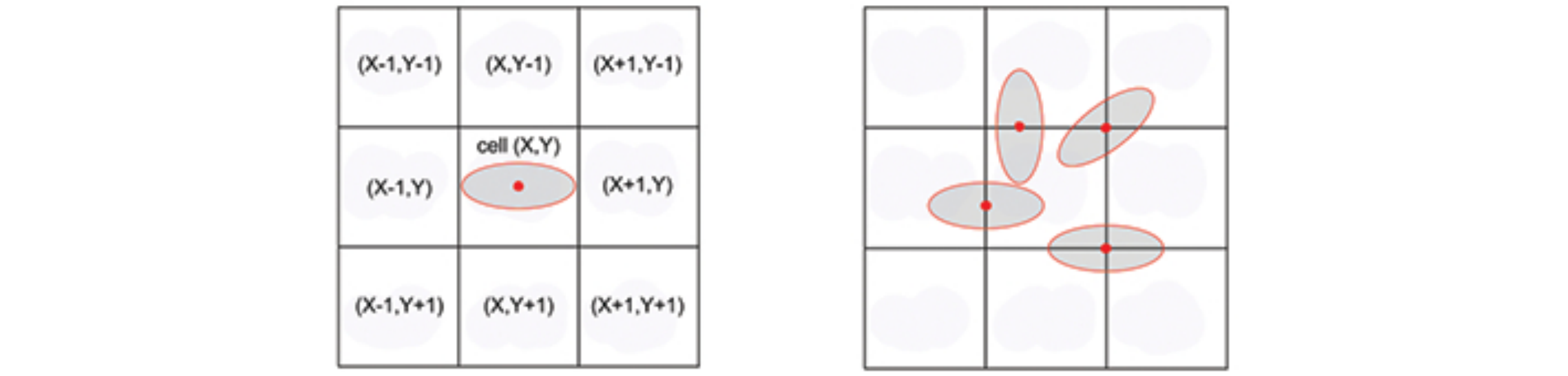}{fig:partitioning}{The $C^{3}$ sky partitioning method. The sky is partitioned in cells whose dimensions are determined by the maximum dimension that the matching regions can assume (panel \emph{b}). Each source of the second catalogue is assigned to a cell; a source and an ellipse referred to a first catalogue object can match only if the source lies within the nine cells surrounding the object (panel \emph{b}).}

\section{A $C^3$ customized version: ClumpPopulator}

\articlefigure[width=.48\textwidth]{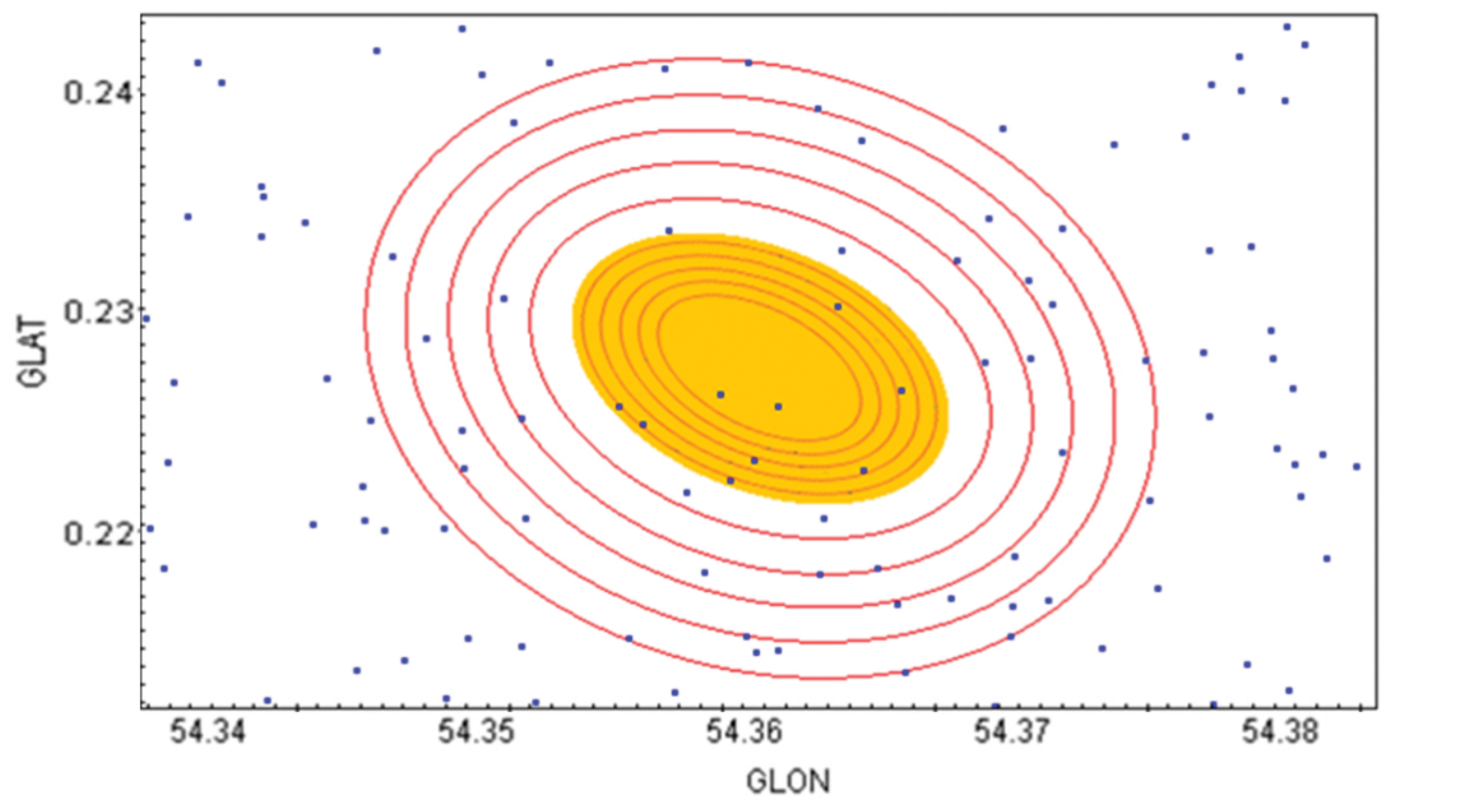}{fig:cp}{ClumpPopulator is able to perform positional association also for a number of additional ellipses, concentric to the basic clump ellipse and with gradually increasing and/or decreasing dimensions.}

In the FP7 project ViaLactea, \citep{molinaro2016}, a customized $C^3$ version, named \textit{ClumpPopulator}, has been designed to find positional associations between sources detected at different resolutions. This software has been calibrated to be applied to the catalogue of clumps produced by Hi-GAL \citep{molinari2016} and a cross-matched catalogue of sources from UKIDSS \citep{lucas2008} and GLIMPSE \citep{churchwell2009} surveys, although it can work on arbitrary source catalogues. In addition to the cross-match based on the catalogue parameters (FWHMs and position angles at a given resolution), the association is extended to a user-defined number of additional ellipses, concentric to the basic clump ellipse and with gradually increasing and/or decreasing dimensions. Moreover, a set of routines have been provided to remove intersecting clumps from the results, to compare stellar surface density inside and outside the clumps and to produce a catalogue containing only the sources associated to clumps.

\section{Conclusions}\label{sect:conclusion}
In this paper we have introduced $C^{3}$, a new scalable tool to cross-match astronomical datasets. It is a Python script, based on the multi-core parallel processing paradigm and on a sky partitioning algorithm in order to ensure high-performance capability, and designed to provide the maximum flexibility to the end users in terms of choice about catalogue properties (I/O formats and coordinates systems), shape and size of matching area and cross-matching type. Despite the general purpose of the tool, it is easy to configure, by compiling a single configuration file, and to execute as a stand-alone process or integrated within any generic data reduction/analysis pipeline. It is also possible to tailor the tool configuration to the features of the hosting machine, by properly setting the number of concurrent processes and the resolution of sky partitioning.


A test campaign, done on real public data, has been performed both to scientifically validate the $C^{3}$ tool, showing a perfect agreement with other publicly available tools, and to compare its computing time efficiency with other cross-matching applications, revealing the full comparable performance, in particular when input catalogues increase their size and dimensions.

The $C^{3}$ tool and the user guide are available at the page \url{http://dame.dsf.unina.it/c3.html}.


\acknowledgements This work was financed by the 7th European Framework Programme for Research Grant FP7-SPACE-2013-1, \textit{ViaLactea - The Milky Way as a Star Formation Engine}.

\bibliography{P1-25}  

\end{document}